\documentclass[sigconf,nonacm,balance=false]{acmart}

\usepackage[utf8]{inputenc}
\usepackage{amsmath}
\usepackage{textcomp}
\usepackage{graphicx}
\usepackage{array}
\usepackage{booktabs}
\usepackage{tabularx}
\usepackage{float}
\usepackage{textcase}

\setcitestyle{authoryear,round,semicolon,aysep={,},yysep={,}}

\renewcommand{\arraystretch}{1.12}
\newcolumntype{Y}{>{\raggedright\arraybackslash}X}


\begin{document}

\title{Cnuas: A Software-Defined AI/HPC Rack-scale Emulation Platform and
Hyperscale Data Center Facility Twin}
\author{{\renewcommand{\arraystretch}{0.97}\begin{tabular}{@{}c@{\hspace{2.5em}}c@{\hspace{2.5em}}c@{}}
\begin{tabular}[t]{@{}c@{}}
Weqaar Janjua\textsuperscript{1\char44 2,*} \\
{\small\textsuperscript{1}University of Limerick, Ireland} \\
{\small\textsuperscript{2}Packet Five Networks Ltd., Ireland} \\
{\small\texttt{janjua.weqaar@ul.ie}} \\
{\small\texttt{weqaar.janjua@packetfive.com}}
\end{tabular} &
\begin{tabular}[t]{@{}c@{}}
Eoin O'Connell \\
{\small University of Limerick, Ireland} \\
{\small\texttt{eoin.oconnell@ul.ie}}
\end{tabular} &
\begin{tabular}[t]{@{}c@{}}
Mihai Penica \\
{\small University of Limerick, Ireland} \\
{\small\texttt{mihai.penica@ul.ie}}
\end{tabular}
\end{tabular}}}
\renewcommand{\authors}{Weqaar Janjua\and Eoin O'Connell\and Mihai Penica}
\renewcommand{\shortauthors}{Janjua et al.}

\begin{abstract}

Modern AI and HPC systems integrate accelerators, high-speed networks, and
management controllers at rack scale. Developing software for this
infrastructure typically requires access to scarce, costly hardware, while
software abstractions can obscure how workloads depend on resources across
servers and accelerators. This paper presents Cnuas, an open-source,
experimental rack-scale emulation platform whose baseline architecture follows
the Open Compute Project (OCP) Open Rack v3 specifications. Through functional
emulation, it supports experimentation, learning and software development
within academic and industrial research and development, rather than matching
the throughput or latency of physical hardware. Its web-based user
interface visualizes racks, devices and their interconnections to help
developers build a system-level mental model of the infrastructure supporting
their workloads.

At its core, CnuasNIC and CnuasSwitch implement a guest-visible remote
direct memory access (RDMA) adapter and a host-resident hybrid software
switch supporting both RoCEv2 and native InfiniBand. The platform also
provides a dedicated AI/ML accelerator (GPU) peer fabric and OpenBMC-based
rack management with executable power supply and battery backup firmware
over RS-485. These components support the study of device, driver,
and firmware interfaces on commodity hosts. The accelerator software stack
remains an early research prototype, and facility modeling with OpenUSD is an
exploratory extension. The paper presents the architecture, interfaces, and
bounded prototype results as a basis for community collaboration across the
core platform and its extensions.
\end{abstract}

\keywords{AI/HPC infrastructure emulation, Open Rack v3, OpenBMC, RDMA,
accelerator emulation, exploratory facility modeling, virtual time}

\maketitle
\hypersetup{pdfauthor={Weqaar Janjua, Eoin O'Connell, Mihai Penica}}
\begingroup
\renewcommand{\thefootnote}{\fnsymbol{footnote}}
\footnotetext[1]{Corresponding author: Weqaar Janjua,
\texttt{janjua.weqaar@ul.ie}.}
\endgroup

\section{Introduction}

AI rack software depends on hardware-specific interfaces. A device driver binds
to a PCI function; a verbs provider uses its queue and doorbell layout.
A subnet manager configures the fabric; rack firmware communicates with
the power shelf over a sideband bus. These
paths are normally exercised together after the hardware is installed.
This delays rack-scale software integration and limits its availability for
continuous integration. Procurement cost and lead time further restrict
access to switches, adapters and accelerators. Shared facilities provide
scheduled allocations rather than continuously researcher-controlled
machines \citep{duplyakin2019,keahey2020}.

Cnuas is intended to broaden access to rack-scale AI/HPC systems for academic
and industrial research and development. Within this context, it supports
experimentation, learning and software development. Its goal is to expose and exercise
device interfaces, protocols and interactions across the software stack,
rather than reproduce the throughput or latency of physical hardware.
This paper evaluates the implemented software functionality.

Cnuas complements physical testbeds and hardware-assisted virtualization,
rather than serving as a performance-equivalent replacement. This distinction
includes single-root I/O virtualization (SR-IOV): supported NICs and GPUs
expose PCI Express (PCIe) virtual functions (VFs), which can be assigned to
virtual machines or made available as VF-backed devices to container workloads
through host drivers. These paths retain a dependency on physical devices;
Cnuas instead provides modifiable software models of the implemented
device interfaces and protocols.

A further motivation for the Open Rack v3 (ORV3) architecture
\citep{ocporv3} is to provide developers and engineers with a concrete mental
model of the infrastructure underlying their workloads. Software interfaces
often conceal the arrangement of compute blades, accelerators, network
adapters, switches and shared power equipment. The Cnuas web interface places
equipment and connections in a visual rack context; Figure~\ref{fig:rack}
shows the underlying reference deployment. The intention is to help users relate application
execution to device placement, data movement, communication paths and shared
infrastructure. This is particularly relevant to distributed AI and HPC
workloads, whose execution can span multiple accelerators and servers. The rack
therefore serves both as an architectural basis for emulation and as a
reference for understanding the system being programmed.

Existing simulators primarily support performance and architectural analysis.
Cycle-level GPU simulators produce timing estimates for a modeled architecture
\citep{bakhoda2009,khairy2020,sun2019}. Network simulators reproduce protocol
behavior and queueing in a discrete-event framework
\citep{riley2010,varga2008}; some also connect real software stacks to simulated
networks \citep{ns3tap}. Cnuas instead focuses on a guest-visible device
boundary spanning the application, provider library, kernel driver and
emulated adapter. For example, \texttt{ibv\_rc\_pingpong} runs through
\texttt{libibverbs} and a kernel driver bound to an enumerated PCI device.
This preserves the software-layer boundaries used with physical adapters.

Cnuas\footnote{Pronounced ``Knoo-us''; the name comes from the Irish
\emph{Gaeilge} word for ``cluster''.} uses functional device emulation for this
purpose. Each guest-facing
adapter or accelerator is a model presented over an emulated PCI bus to a guest
running a stock kernel with a Cnuas driver module and otherwise standard user
space. The software under study therefore crosses the same application, driver
and device boundaries as it would on a physical system. Emulation reproduces
the programming interface and protocol behavior, but it does not by itself
reproduce physical timing; Section~\ref{sec:timing} describes the separate
analytic timing model and its limitations. Cnuas can demonstrate
what a queue pair is, how a subnet manager brings a fabric up, how a collective
is decomposed, and how a kernel reaches an accelerator. Implemented
capabilities and outstanding work are summarized in
Table~\ref{tab:implementation-status}.

Cnuas is being developed as an open-source software product and an extensible
research framework. Its scope exceeds the capabilities currently implemented
and evaluated. Existing components include research prototypes evaluated
through focused functional tests; substantial parts of the wider architecture remain under
development or are planned. The accelerator software stack is at an early
experimental stage. Public release is intended to support collaborative
development of the core platform and its extensions.

Rack-level development also depends on management, power and cooling.
An early facility-modeling prototype explores connecting emulated OpenBMC
power-shelf telemetry to OpenUSD scenes \citep{aousd2025} and simplified load
calculations. Section~\ref{sec:facility} describes this exploratory extension.

The contribution of this paper is threefold.

\begin{enumerate}
  \item A paired RDMA device and fabric implementation: CnuasNIC presents
  a guest PCI adapter with Linux drivers and a standard verbs provider,
  while CnuasSwitch implements a host-resident hybrid soft RoCE and soft
  InfiniBand switch. The pair exposes both protocol paths within the
  emulated system, making device, driver and switch behavior available for
  joint development rather than proxying RDMA to a physical host adapter.
  \item A vertically integrated rack architecture based on Open Rack v3 specifications
  that presents guest-visible RDMA and accelerator devices, separate network
  and accelerator fabrics, real OpenBMC images, executable PSU and BBU firmware
  on an RS-485 multi-drop bus. The architecture also provides an integration
  point for exploratory facility modeling.
  \item A reproducible platform-validation workflow combining component tests
  and firmware demonstrations, with unmet environment requirements reported
  as explicit gates rather than successful results.
\end{enumerate}

The evaluation combines named component tests and recorded firmware
demonstrations. These establish the behavior exercised; broader integration
and application suitability require separate evaluation. Outstanding tests
and unimplemented capabilities are identified throughout the paper.

\begin{table*}[t]
\centering
\caption{Implementation status and evaluation scope of the experimental platform.}
\label{tab:implementation-status}
\begin{tabularx}{\textwidth}{@{}>{\raggedright\arraybackslash}p{0.13\textwidth}
Y Y Y@{}}
\toprule
Component & Existing implementation & Evaluation status & Planned or not established \\
\midrule
RDMA and switch &
Guest PCI adapter, kernel drivers, verbs provider, RoCEv2 and InfiniBand data
paths, top-of-rack switch and integrated subnet manager &
Research implementation; component checks do not establish complete guest
integration, which requires a separate guest-based evaluation &
ETS bandwidth scheduling and further subnet/performance management planned; SA PathRecord,
PMA/\texttt{perfquery}, external OpenSM interoperability and physical
throughput not established \\
CnuasGPU &
Software-emulated accelerator with PCI, driver and runtime interfaces;
standalone host modes &
Early research component; GPU-specific evaluation is outside this paper's scope &
Experimental FPGA design in progress; RTL implementation planned.
CXL planned before UALink; detailed GPU evaluation reserved for a separate paper \\
CnuasLink &
Independent peer-fabric frames, discovery, forwarding and collective payload
transport &
Research implementation with host fabric and client tests &
Further physical-link attachments planned; vendor-fabric compatibility and
performance equivalence not established \\
Rack management &
Real OpenBMC images, QEMU/Renode controllers, six PSU and six BBU firmware
instances, RS-485 Modbus RTU and Redfish telemetry &
Firmware and management-interface tests within emulation &
Additional devices planned; electrical protection and physical
power-hardware equivalence not established \\
Facility modeling &
OpenUSD scene generation, simplified load arithmetic and telemetry bridge &
Early exploratory prototype with software consistency checks &
Scenario validation and reference comparisons required; finished twin,
physical accuracy and operational suitability not established \\
Virtual time &
Deterministic event model, profiles, fitting and error reporting &
Experimental analytic model; no physical calibration dataset &
Target-specific calibration planned; prediction accuracy and live
cross-emulator clock synchronization not established \\
\bottomrule
\end{tabularx}
\end{table*}

\section{Related Systems and Scope}

\subsection{Accelerator simulation}

GPGPU-Sim established detailed simulation of GPU workloads and remains the
reference point for microarchitectural studies \citep{bakhoda2009}. Accel-Sim
extended the approach with a validated framework and broader
front-end support \citep{khairy2020}. MGPUSim addressed multi-accelerator
systems and the interconnect between them \citep{sun2019}, and Multi2Sim
modeled combined CPU and GPU execution \citep{ubal2012}. These tools answer
questions about performance and architecture using traces or emulated
instruction streams. Cnuas prioritizes guest device and driver interfaces
rather than cycle-accurate microarchitectural timing.

MIAOW is the closest open-RTL precedent at the accelerator block level. It
implements an AMD Southern Islands GPGPU in RTL for low-level architecture
experiments and OpenCL compatibility \citep{balasubramanian2015}.
MIAOW and Cnuas expose different experimental surfaces:
open GPU RTL in the former and an integrated software-emulated rack platform in
the latter.

\subsection{Full-system simulation and emulation}

gem5 provides full-system simulation including devices and operating system
boot \citep{binkert2011}, and FireSim accelerates cycle-exact scale-out
simulation using FPGAs in the cloud \citep{karandikar2018}. Both deliver
fidelity that Cnuas does not attempt, at a cost in wall-clock time and, for
FireSim, in infrastructure. QEMU supplies the machine emulation that Cnuas
builds on \citep{bellard2005}, and Renode provides the embedded simulation used
for the management-plane peripherals \citep{renode}. Cnuas contributes device
models rather than a new emulator. The design targets interactive execution of
the integrated rack software stack on a workstation; the current evaluation
is described in Section~\ref{sec:validation}.

\subsection{Shared testbeds}

Emulab pioneered configurable network experimentation \citep{white2002}, and
CloudLab and Chameleon provide bare-metal, reconfigurable resources to the
research community \citep{duplyakin2019,keahey2020}. These support physical
measurements but provide allocated rather than continuously
researcher-controlled access. Cnuas supports locally controlled interface and
integration work; physical testbeds remain necessary for hardware measurements.

There is a second distinction beyond availability. Access to a physical
AI or HPC testbed ordinarily permits experiments above the interfaces exposed
by its installed accelerators; it does not give a user a modifiable
implementation of proprietary accelerator RTL, internal firmware, math blocks,
or fabric logic. Cnuas trades physical fidelity for a vertically open research
surface in which a change can cross the device model, instruction set,
compiler, kernel driver, runtime, libraries, fabrics and rack management,
with facility modeling as an exploratory extension. This provides a basis
for cross-layer co-design questions that a
fixed cluster cannot, while the physical testbed remains necessary for timing,
power, manufacturability, and compatibility measurements.

\subsection{Software RDMA and fabric emulation}

The Linux SoftRoCE driver implements the RoCE transport in software over an
ordinary Ethernet interface \citep{rxe}, and SoftiWARP does the same for iWARP
\citep{metzler2010}. Both make verbs programming possible without an RDMA
adapter. They terminate the transport in the host kernel rather than in an
emulated PCI adapter and do not themselves model a switch or fabric.
Host transport mechanisms, including TCP congestion control for SoftiWARP,
are distinct from switch-level forwarding and fabric management. CnuasNIC is an
emulated adapter behind a PCI function with its own kernel drivers and verbs
provider. Its connection to CnuasSwitch exposes forwarding tables, subnet
management and management datagrams for modification and study.

QEMU's \texttt{pvrdma} is a close precedent for presenting RDMA behind a PCI
function. It
implemented VMware's paravirtual RDMA interface, registering base address
registers and MSI-X vectors and binding the in-tree \texttt{vmw\_pvrdma} driver
in the guest \citep{pvrdma}. It proxied the transport to the host:
QEMU's documentation describes it as acting ``like a proxy between the Guest
Driver and the host ibdevice interface'', mapping each guest queue pair and
completion queue onto a resource requested from a host RDMA device, which could
itself be SoftRoCE. The device model did not implement a fabric or switch;
InfiniBand management traffic needed an external
multiplexer, and the device was deprecated in QEMU 8.2 and removed in 9.1.
Cnuas terminates the transport inside the emulated devices and carries it over
an emulated fabric, so the protocol behavior under study belongs to the model
rather than to a host stack beneath it.

\subsection{Single-machine emulation of a cluster}

Phantora studies distributed training on a single machine by intercepting
accelerator and collective-library calls \citep{qin2026}. It replaces behavior
below those calls with a model for machine-learning system performance
estimation. Cnuas instead retains the driver, provider, runtime and application
above its emulated device boundary. Its current timing model produces analytic
estimates, not validated predictions of physical-system performance.
Phantora therefore addresses training-time estimation, while Cnuas emphasizes
device-interface and cross-layer integration behavior.

\subsection{Collective communication and programming models}

The accelerator software stack follows the shape established by CUDA
\citep{nickolls2008}, by collective libraries such as NCCL \citep{nccl}, and by
partitioned global address space messaging as standardized in OpenSHMEM
\citep{chapman2010}. ASTRA-sim models distributed training on simulated
platforms and is the closest work in intent at the collective layer
\citep{rashidi2020}. Cnuas additionally provides a callable collective library
whose payloads traverse its peer fabric.

\subsection{Facility modeling}

Warehouse-scale computing established the discipline of reasoning about a
data center as one machine, including its power and cooling
\citep{barroso2018}, and power usage effectiveness is the standard efficiency
metric \citep{iso30134}. Facility digital twins are established practice in
industry and in research, and OpenDT is a recent example that combines live
telemetry with simulation to study data center performance and sustainability
\citep{opendt}. Such twins are normally driven by traces or by telemetry from a
physical site. Cnuas explores the use of emulated rack-firmware telemetry as an
input to a campus scene. This early prototype is not evaluated as a substitute
for a validated facility twin such as those studied in this literature.

\subsection{The gap}

Prior work provides mature or specialized artifacts at individual layers, and
the layers are usually studied apart. The literature reviewed for this paper
shows that the systems differ principally in where they place their modeling
boundary and which research questions they address, as summarized in
Table~\ref{tab:prior-systems}. No prior system identified in this review
integrates guest-visible accelerator and RDMA devices, independent network and
accelerator fabrics, executable Open Rack v3 management and power-shelf
firmware, together with an extensible path toward facility modeling.
Cnuas investigates this combination with an emulation boundary below ordinary
guest software and above the modeled devices. Its contribution is the
experimental rack-scale product, its integration interfaces and bounded
prototype results.

\begin{table*}[p]
\centering
\caption{Scope comparison with representative systems. ``Physical'' means the
result comes from allocated hardware rather than a timing model.}
\label{tab:prior-systems}
\begin{tabularx}{\textwidth}{@{}>{\raggedright\arraybackslash}p{0.11\textwidth}
>{\raggedright\arraybackslash}p{0.18\textwidth}
>{\raggedright\arraybackslash}p{0.14\textwidth} Y Y@{}}
\toprule
System & Modeling boundary & Guest device and driver & Fabric, rack and facility scope & Research focus \\
\midrule
GPGPU-Sim / Accel-Sim \citep{bakhoda2009,khairy2020} &
GPU instruction and microarchitecture model & No guest PCI driver path &
GPU and modeled interconnect, not rack management or facility &
GPU performance and architecture estimates \\
gem5 \citep{binkert2011} &
Full-system machine model & Model-dependent guest OS and devices &
Machine and modeled interconnect; no Open Rack management or facility twin &
Architecture and full-system simulation \\
FireSim \citep{karandikar2018} &
FPGA-hosted RTL model & Workload runs on modeled hardware &
Cycle-exact scale-out system; no facility twin &
Cycle-exact scale-out simulation \\
SoftRoCE \citep{rxe} &
Host-kernel RDMA transport & No emulated PCI device &
Host network transport; no modeled switch, rack management or facility &
Functional verbs and transport on the host \\
pvrdma \citep{pvrdma} &
Guest PCI proxy to host RDMA & In-tree guest driver &
Host-backed RDMA proxy; no implemented switch, rack or facility &
Guest access to host RDMA \\
Phantora \citep{qin2026} &
Accelerator and collective library interception & No driver or device path &
Modeled accelerator cluster above the library boundary &
Machine-learning system performance estimation \\
CloudLab / Chameleon \citep{duplyakin2019,keahey2020} &
No model at the device boundary & Physical devices and drivers &
Allocated physical testbed; no user-modifiable facility twin &
Physical compatibility and performance \\
Cnuas &
Guest device, fabric and rack-management models; exploratory facility extension &
Guest PCI devices, Cnuas drivers and standard user space &
Two data fabrics, Open Rack v3 management and early facility prototype &
Bounded functional results and analytic virtual time; not product readiness
or validated physical performance \\
\bottomrule
\end{tabularx}
\end{table*}

\section{System Architecture}

\begin{figure*}[t]
\centering
\includegraphics[width=0.96\textwidth]{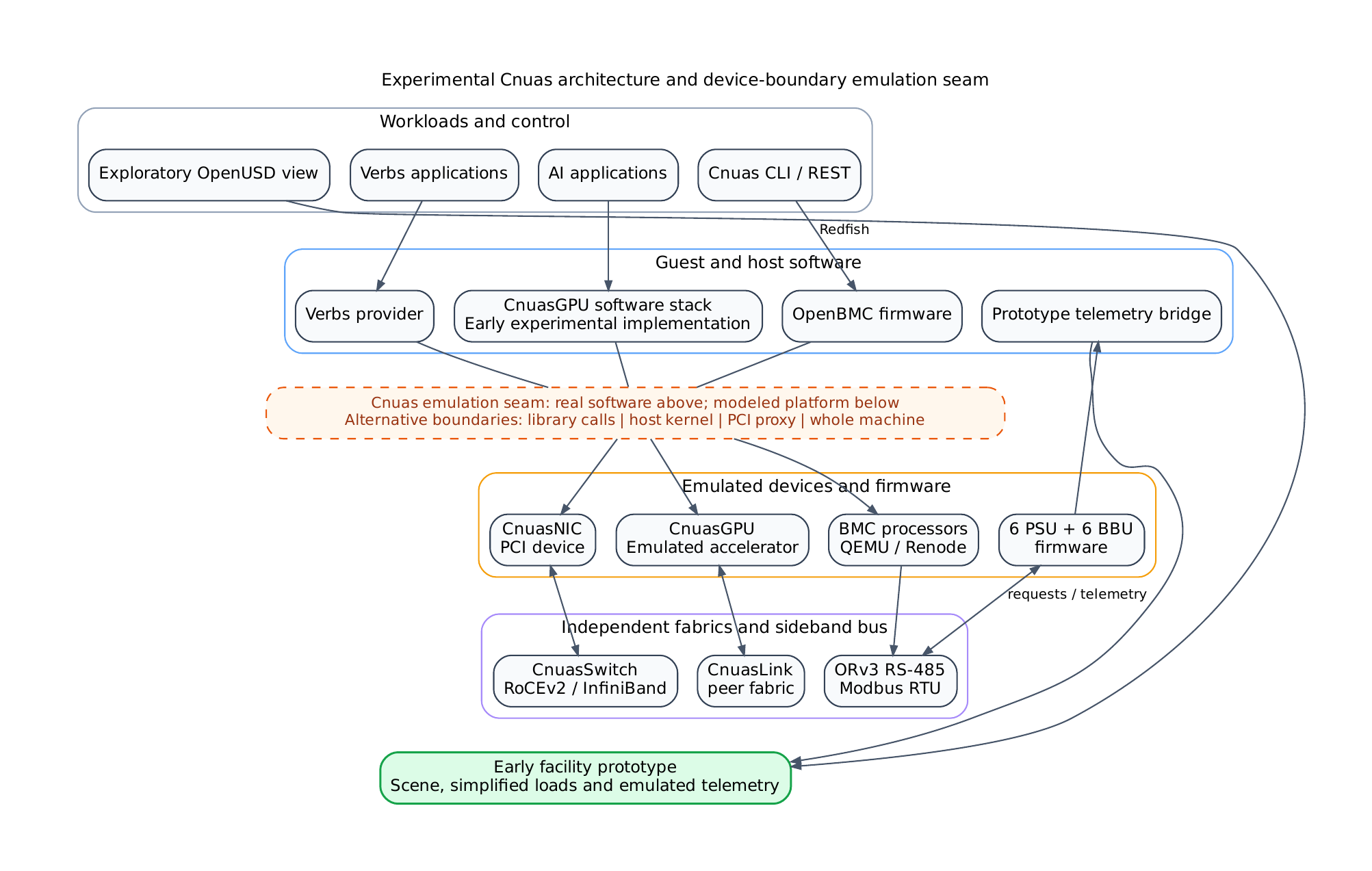}
\Description{Layered Cnuas architecture from workloads and guest software
through the explicit Cnuas device-boundary emulation seam to independent
network and accelerator fabrics, Open Rack v3 management, power-shelf firmware,
telemetry, and an exploratory facility prototype. Alternative library, kernel, proxy and
whole-machine modeling boundaries are identified alongside it.}
\caption{The experimental Cnuas architecture and its emulation seam.
Ordinary software above the seam exercises the modeled platform below it.
The accelerator stack is early-stage and the facility extension exploratory.
Representative alternative boundaries correspond to the systems compared in
Table~\ref{tab:prior-systems}.}
\label{fig:architecture}
\end{figure*}

\subsection{Deployment model}

\begin{figure*}[t]
\centering
\includegraphics[width=0.90\textwidth]{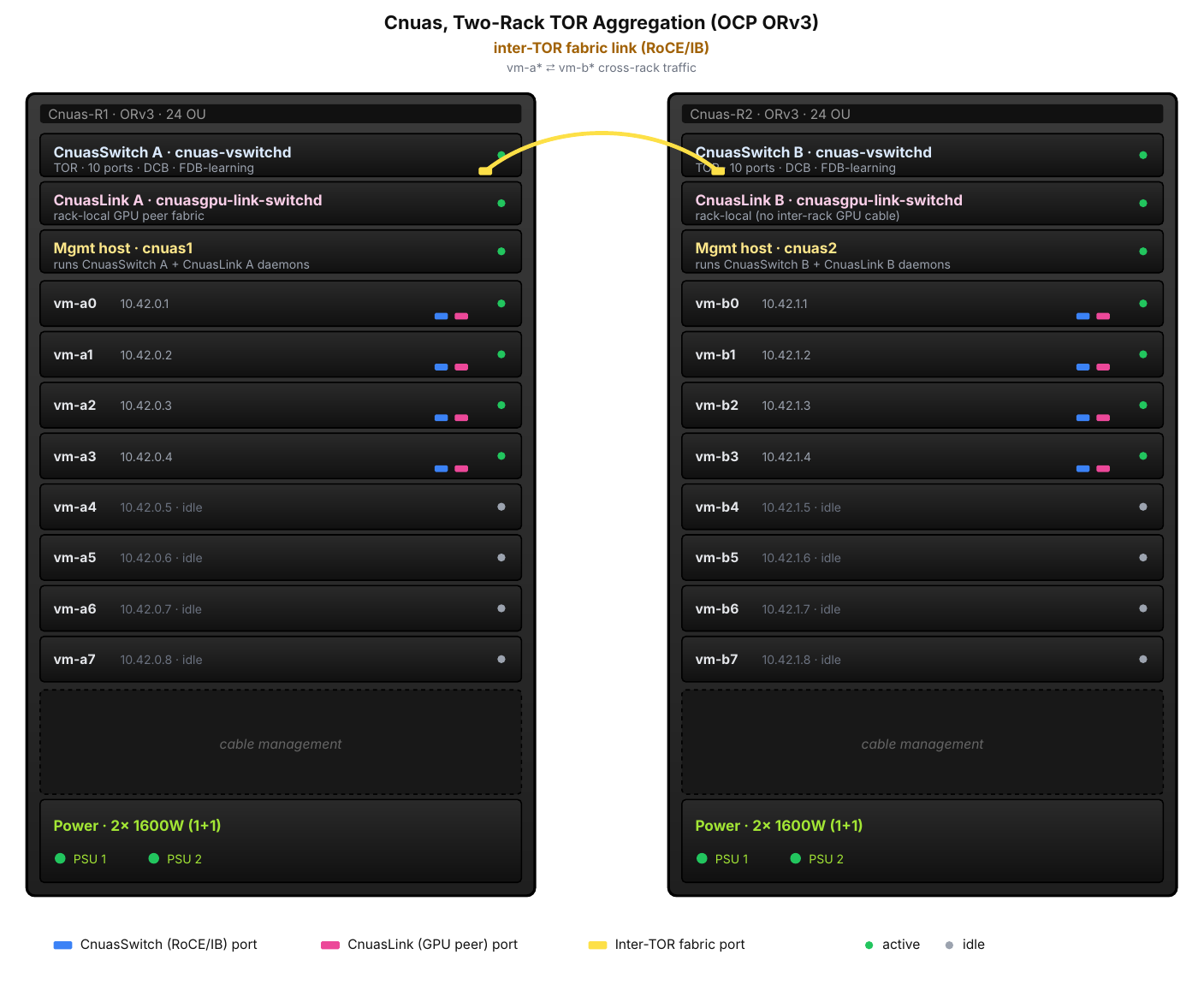}
\Description{Two Open Rack v3 racks, each containing paired top-of-rack
switches, a management host and eight compute blades, connected through the
RDMA, accelerator and management fabrics.}
\caption{The Cnuas reference deployment, two racks modeled on the Open Compute
Project Open Rack v3. Each rack carries two 1-OU top-of-rack switches, one for
the RoCE and InfiniBand fabric and one for the accelerator fabric, a management
host, and eight 2-OU blades. Every blade is a guest with an emulated RDMA
adapter and one or more emulated accelerators; the scope of the current
evaluation is described in Section~\ref{sec:validation}.}
\label{fig:rack}
\end{figure*}

Figure~\ref{fig:architecture} shows the architecture and integration interfaces,
not uniform maturity across its components. The reference deployment, shown
in Figure~\ref{fig:rack}, follows
the Open Compute Project Open Rack v3 layout \citep{ocporv3}.
Equipment height is expressed in Open Rack units (OU). Each rack holds two
1-OU top-of-rack switches, a management host running the daemons, and eight
2-OU blades
acting as compute nodes. Each blade is a guest virtual machine with an emulated
RDMA adapter and one or more emulated accelerators. Two racks are modeled in
the present tree, connected through an inter-rack fabric link.

The host requirement is an x86-64 machine with AVX2, 16~GB of memory as a
minimum with 64~GB recommended, and KVM enabled. AVX-512 is used when present
for the accelerator compute backends. These requirements apply to the
KVM-accelerated x86-64 rack deployment; other host architectures have not been
validated.

\subsection{Platform component families}

\begin{table*}[tp]
\centering
\caption{Platform component families and roles. See Table~\ref{tab:implementation-status}
for implementation and maturity.}
\label{tab:devices}
\begin{tabularx}{\textwidth}{@{}>{\raggedright\arraybackslash}p{0.16\textwidth} Y@{}}
\toprule
Component & Function \\
\midrule
CnuasNIC & Guest PCI RDMA adapter supporting RoCEv2 and native InfiniBand,
with Linux network/RDMA drivers and a standard verbs provider \\
CnuasSwitch & Host-resident, ten-port hybrid soft RoCE and soft InfiniBand
switch; Ethernet forwarding, selected DCB mechanisms and integrated subnet management \\
CnuasGPU & Experimental software-emulated accelerator with device memory,
guest PCI attachment, and driver and runtime interfaces \\
CnuasLink & Eight-port accelerator-to-accelerator fabric, modeling a peer
interconnect distinct from the network fabric \\
CnuasBMC & OpenBMC-based sled, switch and rack management, connected to
emulated PSU and BBU firmware over an RS-485 Open Rack v3 power shelf \\
Facility prototype & Exploratory OpenUSD scene generation, simplified load
calculations and connections to emulated CnuasBMC telemetry \\
\bottomrule
\end{tabularx}
\end{table*}

The first four families implement the compute and communication data planes
described below. CnuasBMC addresses rack management
(Section~\ref{sec:bmc}); the early facility extension is discussed separately
in Section~\ref{sec:facility}.

\subsubsection{CnuasNIC}
\label{sec:cnuasnic}

CnuasNIC is a functional, software-emulated RDMA network adapter that speaks
both RoCEv2 and native InfiniBand to the CnuasSwitch fabric. The
\texttt{cnuas-vnic} QEMU PCIe device exposes memory-mapped control registers
and interrupts to two guest Linux modules: \texttt{cnuas\_net} supplies the
network-device interface and \texttt{cnuas\_ib} registers the RDMA device.
A provider implements the rdma-core verbs application binary interface
(ABI), allowing applications to use standard \texttt{libibverbs} calls
through the Linux \texttt{ib\_uverbs} interface.

Implemented operations include send and receive, RDMA read and write,
and atomic compare-and-swap and fetch-and-add. The adapter supports
reliable connected (RC) and unreliable datagram (UD) queue pairs,
shared receive queues and InfiniBand multicast.
Multi-packet transfers honor the negotiated path maximum transmission unit.
The invariant cyclic redundancy check (ICRC) is computed on transmission and
validated on reception. Management queue pairs QP0 and QP1 support subnet
management and general services, including management-datagram processing
and the subnet management agent. Memory registration covers host memory and
an experimental DMA-BUF import interface for peer-memory integration.

RoCEv2 carries RDMA packets over UDP/IPv4 and Ethernet, using UDP destination
port 4791. The native InfiniBand path instead carries a local routing header
(LRH), base transport header (BTH) and InfiniBand opcodes, preserving its own
addressing and management behavior without Ethernet/IP encapsulation.
These are two protocol paths in the adapter, not alternative labels for the
same Ethernet transport.

The current CnuasNIC endpoint requires the emulated PCI function in a guest.
Building its provider outside the rdma-core source tree does not remove
the kernel-driver dependency. Its contribution is the modifiable device,
driver and provider path connected to the emulated switch; the recorded
integration coverage and remaining guest-environment gates are distinguished
in Section~\ref{sec:validation}.

\subsubsection{CnuasSwitch}
\label{sec:cnuasswitch}

CnuasSwitch is a hybrid soft RoCE and soft InfiniBand RDMA switch. Its
\texttt{cnuas-vswitchd} daemon runs directly on the host and presents ten
logical ports: eight fabric ports, an Ethernet uplink and a
console/observability port. One daemon serves both link layers, with
Ethernet-only, InfiniBand-only or automatic classification selectable per
fabric port.

The Ethernet path models MAC-frame forwarding in the IEEE 802.3 Ethernet
family \citep{ieee8023}, including VLAN-aware source-address learning and
destination lookup in a forwarding database (FDB). It carries ordinary
Ethernet traffic as well as RoCEv2 packets. The native InfiniBand pipeline
forwards LRH/BTH frames using destination local identifiers and a linear
forwarding table (LFT), retaining a separate addressing and management path.

Selected data center bridging (DCB) mechanisms are modeled for the Ethernet
path: priority-based flow-control pause handling (PFC, IEEE 802.1Qbb)
\citep{ieeepfc} and configuration of enhanced transmission selection
(ETS, IEEE 802.1Qaz) \citep{ieeeets}. ETS traffic-class mappings and
bandwidth settings are exposed through management; bandwidth-enforcing
scheduling remains development work. IPv4 explicit congestion notification
(ECN) marking follows the field semantics of IETF RFC 3168 \citep{rfc3168};
ECN is distinct from the IEEE DCB amendments. The models cover these selected
protocol mechanisms, not complete IEEE conformance or physical-link behavior.

The integrated subnet manager and subnet management agent support subnet
sweeps, directed-route management packets, local-identifier assignment and
route-table programming. This brings up the modeled InfiniBand fabric
without a separate OpenSM process. A JSON management interface exposes port,
forwarding, flow-control and subnet-manager state, with telemetry for
observability.

Fabric ports use UNIX-domain \texttt{SOCK\_SEQPACKET} sockets. QEMU devices
and host processes implementing the port framing can therefore attach to
the same switch. Host-only clients support protocol and forwarding
experiments without virtual machines; they do not by themselves expose a
Linux verbs device. Together with CnuasNIC, this separates host-resident
fabric development from guest device/driver integration while preserving
both RoCEv2 and native InfiniBand paths.

\subsubsection{CnuasGPU}
\label{sec:cnuasgpu}

CnuasGPU supplies the experimental accelerator component of the rack.
It presents a software-emulated PCI device with device memory. In the guest,
the Linux driver exposes a character device, \texttt{/dev/cnuasgpuN}, where
$N$ is the device index. CnuasDev provides the device-access interface, and
CnuasRT supplies the application runtime above it. Its accelerator-peer
connection uses CnuasLink, separately from the CnuasNIC network path.
Standalone Soft-GPU modes retain these application interfaces on the host CPU
without QEMU or a guest. The \texttt{cnuasgpu\_host} kernel module exposes
\texttt{/dev/cnuasgpu\_hostN}, backed by kernel-owned host memory, for
applications needing a Linux device node. The in-process mode instead uses
an arena in the calling process and requires neither a kernel module nor
a device node. Default discovery prefers the guest PCI device, then the
host character device, and finally the in-process mode.

Experimental FPGA design work is in progress; RTL implementation is planned.
A separate CnuasGPU paper is planned to describe its architecture, software
stack, numerical validation and performance evaluation.

Compute Express Link (CXL) integration is planned ahead of UALink.
CXL specifies cache-coherent connectivity among processors, memory and
accelerators \citep{cxl}. The planned Cnuas work addresses host/device
memory-interconnect experiments, beginning with selection of a specification
revision and device profile, assessment of QEMU and Linux support, and a
reproducible validation path.

UALink remains a later accelerator-to-accelerator attachment. This ordering is
a development priority; the two interconnects serve different roles.
Both remain planned extensions. PCIe provides the existing guest device
attachment, while CnuasLink carries the current modeled accelerator-peer traffic.

\begin{figure*}[t]
\centering
\includegraphics[width=0.86\textwidth]{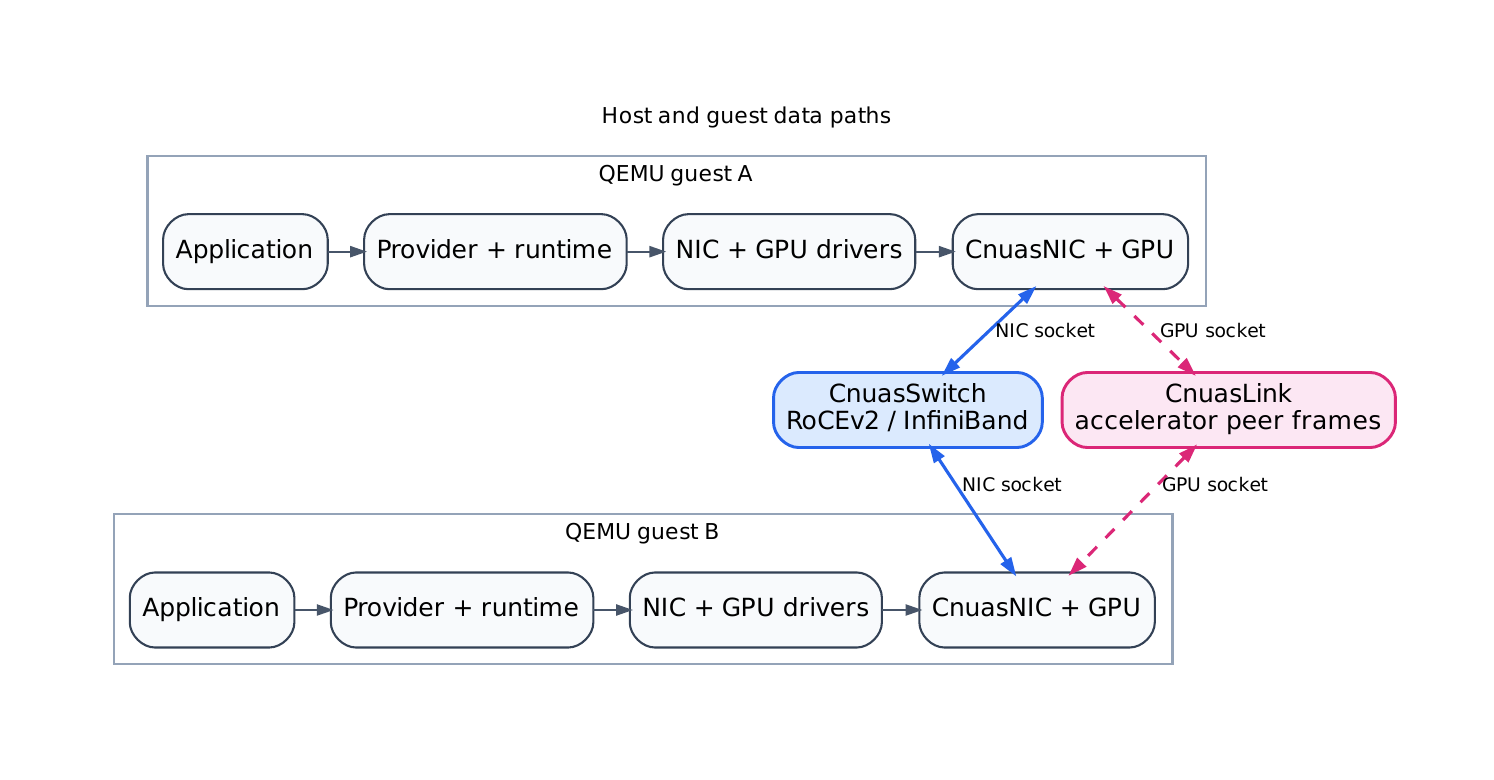}
\Description{Two QEMU guests, each containing applications, provider and
runtime libraries, kernel drivers, CnuasNIC and CnuasGPU devices. Their NICs
connect to CnuasSwitch while their GPUs connect independently to CnuasLink.}
\caption{Host and guest data paths. Applications cross ordinary userspace,
kernel-driver and emulated-device boundaries. RoCEv2 and InfiniBand traffic
uses CnuasSwitch, while accelerator peer traffic uses the independent
CnuasLink fabric.}
\label{fig:dual-fabric}
\end{figure*}

\subsubsection{CnuasLink}

Accelerator-to-accelerator traffic uses a separate fabric with its own switch
daemon, frame format, forwarding database, and discovery protocol, modeling the
peer interconnect of a multi-accelerator node rather than reusing the network
fabric. Keeping the two fabrics apart reflects how such systems are actually
built and allows their control and data paths to be changed independently.

\subsection{Control plane}

One command surface drives the switch fabric, the accelerator fabric, the
accelerators, the adapters, and the lifecycle of the guests, and the same
service layer is exposed as a REST interface for automation. Component daemons
are reached over JSON on UNIX sockets. The uniformity avoids a collection of
unrelated component tools and gives automation one control surface.

A web-based frontend complements these programmatic interfaces with rack and
switch-chassis views. It displays equipment placement, inter-rack links and
switch-port status, giving users a visual context for inspecting the modeled
system.

\section{Exploratory Facility Modeling of the Hyperscale Data Center}
\label{sec:facility}

Facility modeling with OpenUSD \citep{aousd2025} and NVIDIA Isaac Sim
\citep{nvidiaisaac2026} is an early-stage exploratory direction. The initial
Cnuas Facility Twin prototype investigates scene generation, simplified load
calculations and connections to emulated rack telemetry.

\begin{figure*}[t]
\centering
\includegraphics[width=0.82\textwidth]{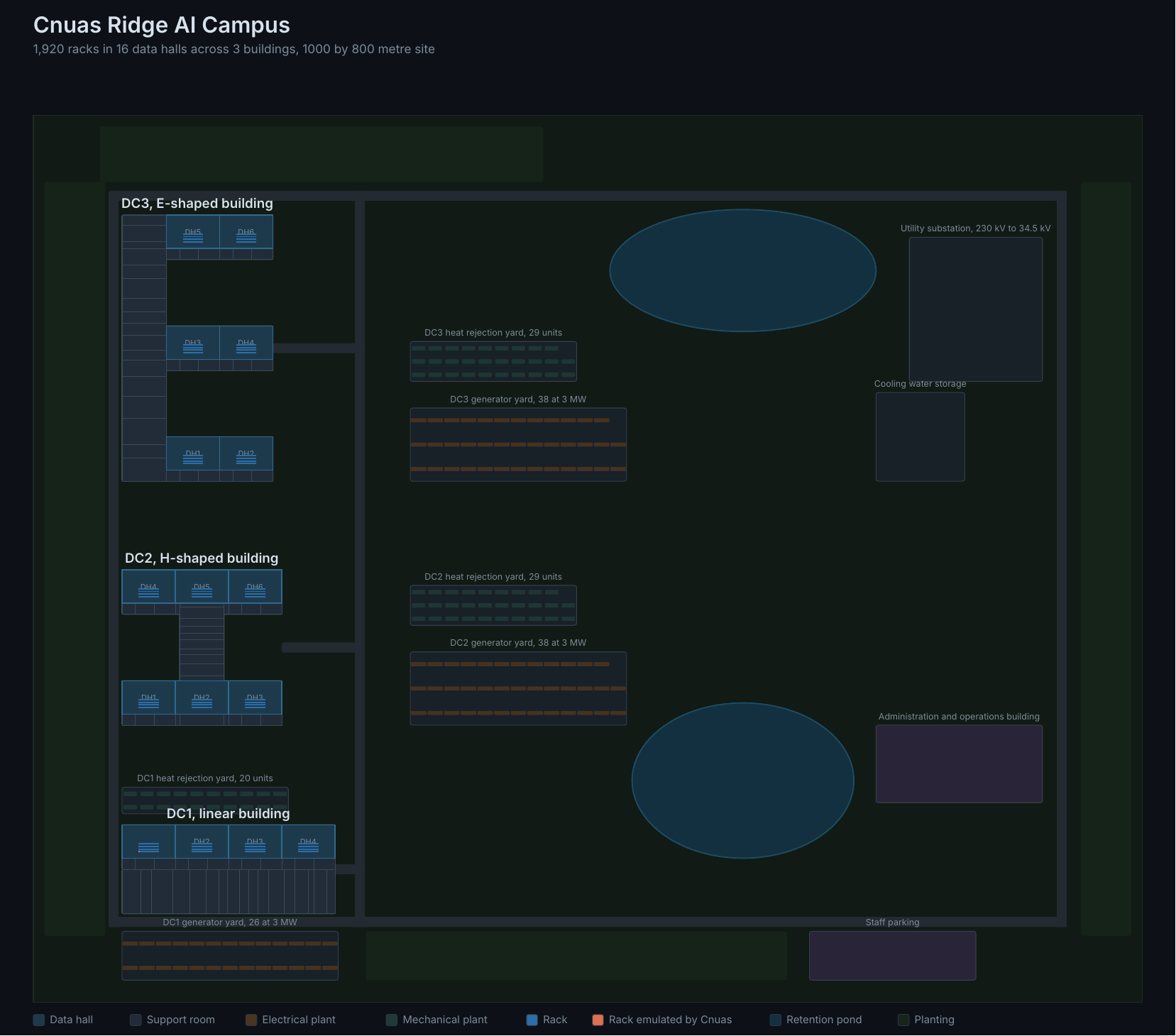}
\Description{Illustrative plan from the early Cnuas facility prototype showing three
data-center buildings, sixteen data halls, utility and cooling plant,
generator yards, roads, offices, landscaped areas and ponds.}
\caption{Illustrative campus layout from the exploratory facility prototype.
Equipment placement and plant representations are provisional.}
\label{fig:campus}
\end{figure*}

The prototype describes a campus declaratively, from buildings and data halls
to rack rows, support rooms, plant and landscaped areas. It generates OpenUSD
geometry for exploration in Isaac Sim. Figure~\ref{fig:campus} illustrates
this scene-generation work.

The load arithmetic sums configured rack ratings and applies an assumed power
usage effectiveness (PUE) multiplier. PUE is formally an energy ratio over a
common measurement period \citep{iso30134}; its use here as a power multiplier
is a simplifying scenario assumption. Heat-load and plant-count calculations
likewise use simplified assumptions.

An experimental bridge reads the emulated Open Rack v3 shelf over Modbus RTU
and writes PSU measurements and BBU state into OpenUSD attributes and time
samples. This explores a connection between firmware and a scene.

Substantial scenario-specific validation remains necessary. This includes
checking geometry and equipment assumptions, comparing electrical and thermal
behavior with appropriate reference data, exercising telemetry timing and
fault conditions, and evaluating calibrated models on held-out scenarios.
Existing software tests check specified arithmetic, configuration handling,
scene structure and telemetry processing. Engineering or operational use
requires physical and scenario validation beyond the present work.
The early prototype will be made available for research and community
development, with further development and validation continuing after release.

\section{\NoCaseChange{CnuasBMC} Rack and Platform Management}
\label{sec:bmc}

Rack management connects the compute nodes to the power architecture and
sideband buses of Open Rack v3 \citep{ocporv3}. CnuasBMC reproduces these
interfaces so that power-control and telemetry workflows can be exercised with
the rest of the emulated rack.

\subsection{The OpenBMC port}

CnuasBMC is the Cnuas OpenBMC port \citep{openbmc}. Its
\texttt{meta-cnuas} Yocto layer defines three machines: sled, top-of-rack
switch and rack management controller. Each has a Cnuas device tree and branded
firmware image. The images boot a real Linux kernel and bootloader on an
emulated ASPEED AST2600. The userspace is OpenBMC rather than a replacement:
sensors are published on D-Bus and served by bmcweb over Redfish, inventory
comes from Entity Manager, host power passes through the normal power-control
service, the managed console uses Serial over LAN, and IPMI answers over the
network. Redfish and IPMI power requests drive GPIO transitions that control a
separate QEMU blade and receive power-good and POST state in return.

\subsection{RS-485 multi-drop power-shelf bus}

The emulated RS-485 segment preserves a shared-bus topology: one master
addresses multiple peripheral nodes, every node receives each request, and only
the addressed node replies. A Cnuas Renode plugin models that shared wire and
exposes one TCP master endpoint to the AST2600 UART. The power peripherals run
as Cortex-M machines in Renode \citep{renode}, while QEMU runs OpenBMC. This
split preserves the multi-drop topology that a QEMU serial character device
alone cannot express.

The emulated shelf contains six PSU firmware instances at addresses \texttt{0xC0}
through \texttt{0xC5} and six BBU firmware instances at \texttt{0x40} through
\texttt{0x45}. One portable firmware image uses straps to select role, slot
and address for each of the twelve Cortex-M instances. They speak Modbus
RTU at 19,200 baud using register maps from Meta rackmon
\citep{metarackmon}. Framing,
CRC-16, address filtering, holding-register reads, protected writes, exception
responses, PSU load behavior, mains loss, battery discharge and fault state
all execute in firmware on the addressed node.

The \texttt{cnuas-rackmond} master sweeps the twelve nodes, converts the
registers to engineering units and publishes 56 D-Bus sensor objects for power,
voltage, current, temperature, fan speed and battery state of charge. bmcweb
serves those objects through the unmodified Redfish chassis sensor collection.
The exploratory bridge in Section~\ref{sec:facility} can consume measurements
originating in firmware responding to a Modbus request. This provenance
distinguishes the data from display-only values. Physical accuracy and
facility-scenario validation remain separate research requirements.

\begin{figure*}[t]
\centering
\includegraphics[width=0.64\textwidth]{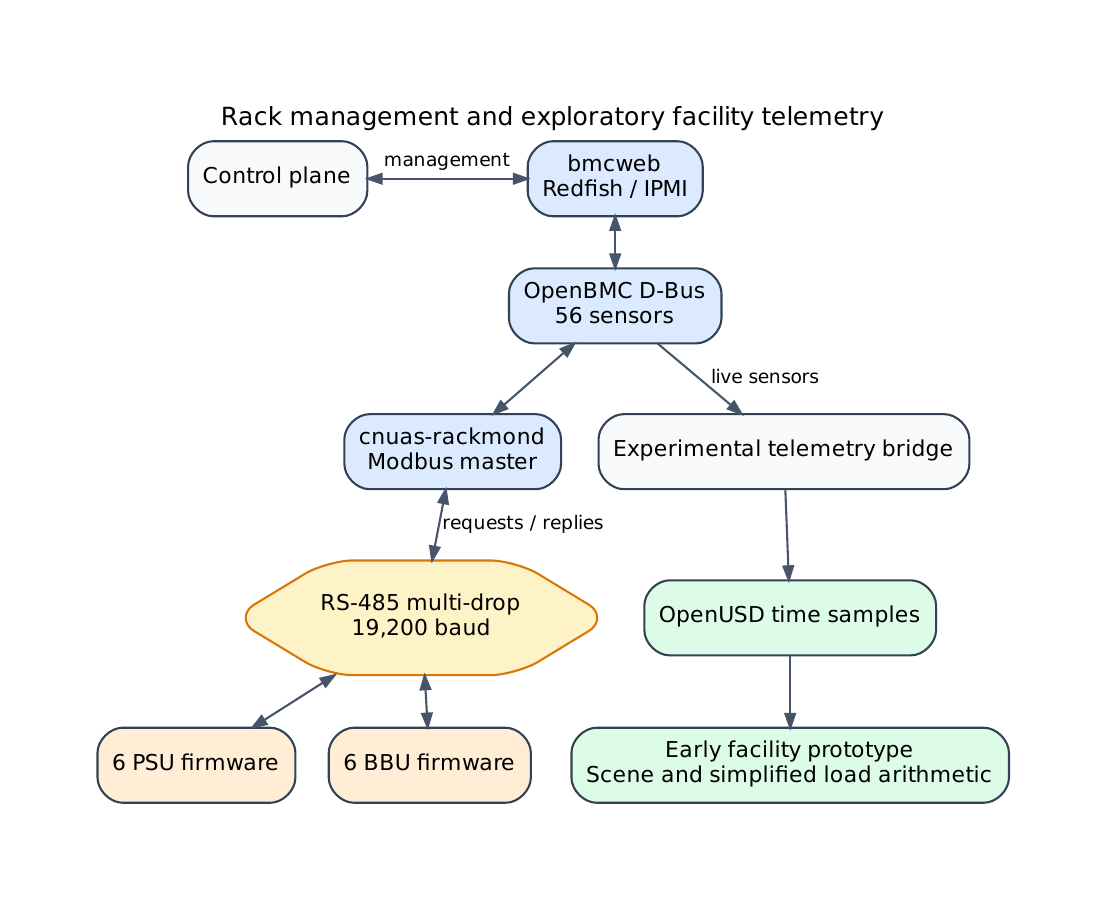}
\Description{Management flow from an operator through bmcweb, OpenBMC D-Bus
and cnuas-rackmond to twelve PSU and BBU firmware instances on RS-485, with
sensor values flowing through an experimental telemetry bridge into an
exploratory OpenUSD facility scene.}
\caption{Rack-management interfaces and the exploratory facility-telemetry
path. Emulated PSU and BBU readings pass through normal OpenBMC sensor services;
the scene connection is an early prototype.}
\label{fig:management}
\end{figure*}

\section{Virtual Time and Calibration Model}
\label{sec:timing}

Timing analysis is a supplementary research capability, separate from
Cnuas's primary role as a functional platform for academic and industrial
research and development.
Functional emulation reproduces interface and protocol behavior.
Physical performance models instead estimate timing. Host wall-clock time
does not provide a reliable mapping between them:
dynamic translation, host scheduling, acceleration mode and workload on the
development machine all affect it. QEMU instruction counting supplies a
deterministic guest clock \citep{bellard2005,qemuicount}, but an instruction count does
not state how long PCIe, DMA, a device queue, accelerator memory, a fabric hop,
an interrupt or contention would take.

Cnuas includes the \emph{Cnuas Calibrated Virtual-time and
Performance Model}. A versioned profile gives each component a fixed cost and
rates for the resources it consumes: instructions, cycles, bytes, packets and
operations. Resources may overlap, in which case the bottleneck term sets the
dynamic service time, or execute serially. An integer-picosecond discrete-event
scheduler assigns work to stable component lanes and records submission, queue,
start and completion times. Repeated inputs produce an identical event trace.
The initial profile covers the guest CPU, PCIe, CnuasNIC, CnuasSwitch,
CnuasGPU compute and memory, CnuasLink, the CnuasBMC RS-485 segment and the
exploratory facility-update cadence.

For overlapping resources the uncalibrated service time follows a
resource-bottleneck, or roofline-style, model \citep{williams2009}:
\begin{equation}
t_a=t_0+\max\left(
\frac{I}{R_I},\frac{C}{f},\frac{8B}{R_B},
\frac{P}{R_P},\frac{O}{R_O}\right).
\label{eq:timing-analytic}
\end{equation}
Here $t_0$ is fixed latency; $I$, $C$, $B$, $P$ and $O$ are respectively
instruction, cycle, byte, packet and operation counts. Rates $R_I$, $f$, $R_B$,
$R_P$ and $R_O$ are measured in instructions/s, cycles/s, bits/s, packets/s and
operations/s, respectively; the factor eight converts bytes to bits.
The maximum is used because
overlapping resources complete when their slowest required resource completes.
For resources configured as serial, the scheduler adds their service times
instead. A target-specific calibration then uses ordinary least squares to fit
the affine correction
\begin{equation}
t_p=\max(0,\alpha t_a+\beta)
\label{eq:timing-affine}
\end{equation}
to training measurements, where $\alpha$ is a scale factor, $\beta$ is a
constant offset and the outer maximum prevents a negative predicted duration.
A separate validation operation reports mean
absolute error, root mean squared error, mean absolute percentage error and
maximum absolute error on held-out observations. The target name and fitted
coefficients become part of the component profile. This design follows the
general requirement that a performance model be validated rather than assumed
\citep{khairy2020,karandikar2018}, while extending one event model across the
network, accelerator, management and facility domains.

The implementation has 46 automated cases covering unit conversion, serial and
overlapping resources, queue contention, parallel lanes, deterministic replay,
strict profile loading, QEMU \texttt{icount} mapping, fitting and held-out
error. These cases cover the listed model behaviors, not
prediction accuracy for a physical target. The shipped values that are not
fixed by protocol are labeled modeling assumptions. No physical calibration
dataset is supplied, so no measured prediction error is reported. The current
implementation evaluates explicit pipelines offline, emits a deterministic
QEMU TCG and \texttt{icount} configuration, and maps supplied instruction
counts. It does not synchronize the live clocks of QEMU, Renode and the host
switch daemons. A fixed conversion such as ten virtual packets per second to
one physical packet per second is therefore not portable or supported by the
present results.

\section{Evaluation}
\label{sec:validation}

\subsection{Methodology}

The evaluation uses a validation matrix to associate each component with
its test coverage and recorded results. A passing test establishes only
the behavior exercised under its recorded configuration.

Tests requiring unavailable devices or environments are reported as skipped,
with the missing requirements recorded. They are not counted as passes.

\subsection{Results}

Table~\ref{tab:test-results} summarizes the platform-related suites selected
from a recorded run on a single workstation. The automated test runner records
host and tool versions, exact commands, exit codes, durations, raw logs,
JUnit records and unavailable environment requirements. It generates CSV,
Markdown and \LaTeX{} summaries from these records and includes SHA-256
checksums for file-integrity verification.

The Python suites cover the control plane, tooling, facility prototype, calibrated
virtual-time model, switch data plane, and end-to-end RDMA paths. CnuasBMC adds
native host checks for its Open Rack v3, rack-controller, and front-panel firmware.

\begin{table}[H]
\centering
\caption{Recorded platform test results. Cases requiring unavailable
devices, guests or lab setups are reported as skipped.}
\label{tab:test-results}
\begin{tabularx}{\columnwidth}{@{}Y r r r@{}}
\toprule
Suite & Cases & Passed & Skipped \\
\midrule
Superproject, switch data plane and end-to-end RDMA & 70 & 45 & 25 \\
Control plane & 52 & 52 & 0 \\
Build, image and guest lifecycle tooling & 123 & 123 & 0 \\
Facility prototype software checks & 66 & 66 & 0 \\
Calibrated virtual-time model & 46 & 46 & 0 \\
Accelerator fabric switch and client & 18 & 18 & 0 \\
\midrule
Total Python & 375 & 350 & 25 \\
\midrule
CnuasBMC Open Rack v3 firmware & 177 & 177 & 0 \\
CnuasBMC rack controller & 45 & 45 & 0 \\
CnuasBMC front-panel firmware & 98 & 98 & 0 \\
\bottomrule
\end{tabularx}
\end{table}

The facility cases check software behavior within the prototype; they are not
physical or scenario validation.

Test counts come from JUnit testcase records or totals emitted by native
test programs. Build-only checks are reported through their commands,
exit status and logs, without testcase counts. Full two-node RDMA testing
requires a guest environment and was not executed in this host run.
The run manifest records its prerequisites and reproduction commands.

\subsection{Lab Testbed Specifications}

The lab testbed comprises two Lenovo ThinkStation P520 workstations.
Each workstation has an Intel Xeon W-2133 CPU at 3.60~GHz with six
physical cores and twelve hardware threads, 128~GB of DDR4 memory and
a 1~TB SSD. Each CPU core has 32~KiB of L1 data cache,
32~KiB of L1 instruction cache and
1~MiB of L2 cache, giving aggregate L1 data, L1 instruction and L2
capacities of 192~KiB, 192~KiB and 6~MiB per workstation. The shared L3
cache is reported as 8.3~MiB. The processor supports AVX-512F,
AVX-512DQ, AVX-512CD, AVX-512BW and AVX-512VL.

Each workstation is equipped with a Mellanox MCX354A-FCCT ConnectX-3 Pro
adapter on a PCIe x8 interface. The adapter provides two QSFP ports
supporting FDR InfiniBand and 40-Gigabit Ethernet.

\section{Limitations}

Table~\ref{tab:implementation-status} summarizes the component-level limitations.
The following limitations govern interpretation of the results.

\textbf{Research maturity.} Cnuas is an experimental software product.
The accelerator stack is early-stage; substantial development across the wider
platform remains in progress or planned. General reliability, usability and
production readiness require broader evaluation.

\textbf{Physical-performance prediction.} Section~\ref{sec:timing}
demonstrates deterministic virtual time, queueing semantics, versioned
profiles, calibration fitting, and held-out error calculation. The supplied
profile has not been fitted to measurements from a named physical target.
Its durations are therefore analytic estimates. Physical-target prediction
accuracy remains to be evaluated.

\textbf{Executable scale.} The repository's reference topology defines two
racks with eight blade slots each. The current recorded host measurement boots one blade, and
the host test run does not include full two-node RDMA execution.
Concurrent execution of all sixteen blades remains to be demonstrated.

\textbf{Facility modeling.} The facility extension is an exploratory prototype.
Scenario-specific reference comparisons, calibration and fault studies remain
necessary to assess physical accuracy and engineering suitability.

\textbf{Advanced InfiniBand management.} The integrated subnet manager assigns
local identifiers and programs the implemented fabric routes. Subnet
Administration PathRecord responses, Performance Management Agent support for
\texttt{perfquery}, and interoperability with an external OpenSM instance are
outside the present validated surface.

\section{Conclusion}

This paper presents Cnuas as an experimental contribution to rack-scale AI/HPC
emulation: an open-source software product under development, with an extensible
architecture, explicit interfaces and bounded prototype results. Existing
device, fabric and rack-management implementations provide a basis for further
work. The primary objective is accessible functional experimentation with
the AI/HPC software stack for academic and industrial research and development,
rather than parity with hardware execution speed.
The accelerator software stack remains early-stage, and facility modeling is an exploratory
extension.

CnuasNIC and CnuasSwitch are central to this contribution: they place the
dual RoCEv2/native-InfiniBand adapter and hybrid software switch within one
modifiable rack environment, while retaining the Linux device, driver and
verbs interfaces used by applications.

The intended development model is collaborative, allowing community
contributions to the Cnuas core, accelerator stack and extensions after public
release. The early facility prototype will be made available in that spirit.
Further work includes full-rack functional integration, characterization of
host resource use and broader reliability testing.

Component-specific follow-on papers are planned to describe four
parts of this architecture in greater depth: CnuasSwitch and its RoCEv2 and
native InfiniBand data paths; CnuasGPU and its accelerator architecture and
software stack, including planned CXL and later UALink integration;
CnuasNIC and its RDMA, InfiniBand, and RoCEv2 device, driver,
and provider implementations; and CnuasLink as the separate accelerator peer
fabric.

Planned communication-software work includes assessing and integrating
Berkeley's UCCL \citep{uccl} over Cnuas's emulated RDMA interfaces, starting
with transport compatibility and bounded transfer and collective-correctness
experiments. UCCL support remains a future integration target, distinct from
the existing CnuasCCL library.

\section*{Acknowledgements}

AI tools assisted in the development and refinement of this work under the
Linux Kernel AI Coding Assistants Policy \citep{linuxcodingassistants}. The lead
author independently verified all outputs and assumes full responsibility for
the final design, analysis, and written content.

\section*{Software and Data Availability}

Project site, documentation, component datasheets and validation matrix:\\
\url{https://cnuas.io}

The project website also provides an illustrative walkthrough of the early
facility prototype. The public source-repository
location will be announced there after release review. The early-stage facility
prototype will be made available for experimental use and community development.
The versioned
upstream contribution artifact is archived on Zenodo at
\url{https://doi.org/10.5281/zenodo.22650437}. It contains the evaluated
component patches and associated provenance, licensing, citation, and
integrity metadata, but not the complete Cnuas source repository or
evaluation dataset.

Upstream submissions remain pending. They comprise the CnuasNIC Linux netdev
and RDMA drivers and UAPI identifier described in
Section~\ref{sec:cnuasnic}; QEMU models for CnuasNIC and CnuasGPU, together
with the ASPEED GPIO and CMIS optical-module support used by CnuasBMC, described
in Sections~\ref{sec:cnuasnic}, \ref{sec:cnuasgpu}, and \ref{sec:bmc}; and the
CnuasBMC OpenBMC machine layer, rack-management service, and Entity Manager
configuration described in Section~\ref{sec:bmc}. Their intended review paths
are the Linux networking and RDMA maintainer trees, QEMU's
\texttt{qemu-devel} process, and OpenBMC Gerrit, respectively. The versioned
artifact carries the evaluated component patches, so their public availability
does not depend on upstream acceptance.

Product and company names used in this paper are the property of their
respective owners and are used for identification only. Mention of a product
or specification does not imply affiliation, endorsement, certification
or conformance.

% === REFERENCES ===
\bibliographystyle{ACM-Reference-Format}

\end{document}